\documentclass[journal]{IEEEtran}

\usepackage{soul,color}
\usepackage{hyperref}

\usepackage{epsfig,epsf,epstopdf}
\graphicspath{ {./Figs/} }
\usepackage[cmex10]{amsmath}
\usepackage{amsfonts}
\usepackage{amssymb}
\usepackage{algorithm}
\usepackage[noend]{algpseudocode}
\makeatletter
\def\BState{\State\hskip-\ALG@thistlm}
\makeatother
\usepackage{cite}
\usepackage{paralist}
\usepackage[dvipsnames]{xcolor}
\usepackage{mathrsfs}
\usepackage{multirow}
\usepackage{subfigure}
\usepackage{graphicx}
\usepackage{caption}
\usepackage[font=small,justification=justified,singlelinecheck=false]{caption}
\usepackage{float}
\ifCLASSINFOpdf
 
\else
 
\fi
\clearpage

\begin{document}

\title{Distributed Resilient Control of DC Microgrids Under Generally Unbounded FDI Attacks}

\author{Yichao Wang, Mohamadamin Rajabinezhad, Omar A. Beg and Shan Zuo
\thanks{Yichao Wang, Mohamadamin Rajabinezhad and Shan Zuo are with the department of electrical and computer engineering, University of Connecticut, CT 06269, USA; Omar A. Beg is with the department of electrical engineering, The University of Texas Permian Basin, TX 79762, USA. (E-mails: yichao.wang@uconn.edu; mohamadamin.rajabinezhad@uconn.edu; shan.zuo@uconn.edu; beg\_o@utpb.edu)

This work has been submitted to the IEEE for possible publication. Copyright may be transferred without notice, after which this version may no longer be accessible.}
}
 
\maketitle

\begin{abstract}

Due to the nature of distributed secondary control paradigm, DC microgrids are prone to malicious cyber-physical attacks, which could be unbounded to maximize their damage. Existing resilient secondary control methods addressing unbounded attacks require that the first time derivatives of cyber-physical attack signals be bounded. The secondary defense strategy presented in this letter relax such a strict constraint by addressing more generally unbounded attack signals and hence, enhance the resilience of DC microgrids in adversarial environments. Rigorous proofs, based on Lyapunov techniques, show that the proposed method guarantees the uniformly ultimately bounded convergence for both voltage regulation and proportional load sharing under generally unbounded attacks. Comparative case studies further validate the enhanced resilience of the proposed attack-resilient control strategy.
\end{abstract}

\begin{IEEEkeywords}
DC microgrids, distributed control, resilient control, unbounded attacks.          
\end{IEEEkeywords}

\section{Introduction}

DC microgrids provide significant advantages due to their compatibility with distributed energy resources, storage units, and predominantly DC-operating modern loads. The hierarchical control stands as the preferred solution for DC microgrids \cite{7}. The droop-based primary control, relying on local measurements, ensures dynamic voltage regulation. Secondary control aiming for global voltage consistency and fair load distribution compensates for voltage anomalies unmanaged by primary control. Distributed secondary control, where each unit operates with a local controller requiring only neighboring data, emerges superior to centralized control, offering improved efficiency, scalability, reliability, and streamlined communication network. However, cyber threats challenge microgrids, stemming from their dependence on digital and communication tools. False data injection (FDI) attacks are of particular concern as they can bypass most attack-detection systems, jeopardizing microgrid operations \cite{14}. While recent advances focus on attack-detection enhancements, many require prompt attack identification and mitigation, a task computationally heavy due to incorporation of non-local communication layer data. Consequently, there is a discernible shift towards resilient control methods as countermeasures.

Existing methods generally consider bounded attacks \cite{liu2021resilient}. In practice, the attacks may be unbounded as in \cite{zhou2023distributed,am29}, presenting greater challenges and potentially leading to more extensive damage. However the solutions in aforementioned papers require the first time derivatives of the attack signals be bounded. The inherent unpredictability of cyber-physical attacks underscores the need for defense strategies resilient to generally unbounded attacks, essential for DC microgrids protection and security. This letter presents a cyber-physical defense strategy for DC microgrids targeting generally unbounded attacks. The contributions are: i) the proposed cyber-physical defense strategy enhances the self-resilience of DC microgrids by addressing generally unknown and unbounded attacks, which cannot be handled by existing solutions \cite{zhou2023distributed,am29}, where the attack signals are strictly required to have bounded first time derivatives; ii) the proposed solution is fully distributed, requiring no global information and hence, is scalable; and iii) using Lyapunov techniques, rigorous mathematical proof is provided for achieving uniformly ultimately bounded (UUB) stability for voltage regulation and proportional load sharing.



\section{Attack-Resilient Controller Design}
Consider a communication network with $N$ converters and one leader node. The connections among local converters are represented by ${\mathscr{G}_f} = (\mathcal{W},\mathcal{E},\mathcal{A})$ with a nodes set $\mathcal{W}$, an edge set $\mathcal{E} \subset \mathcal{W} \times \mathcal{W}$, and an adjacency matrix ${\cal A} = [{a_{ij}}]$. A graph edge, indicating the information flow from converter $j$ to converter $i$, is shown by $({w_j},{w_i})$, with the weight of ${a_{ij}}$. Node $j$ is considered as the neighbor of node $i$ if $({w_j},{w_i}) \in \mathcal{E}$. The set of neighbors of node $i$ is denoted as ${\mathcal{N}_i} = \left\{ {\left. j \right|({w_j},{w_i}) \in \mathcal{E}} \right\}$. The in-degree matrix is $\mathcal{D} = \operatorname{diag}({d_i})$ with ${d_i} = \sum\nolimits_{j \in {\mathcal{N}_i}} {{a_{ij}}} $. $\mathcal{L} = \mathcal{D} - \mathcal{A}$ represents the Laplacian matrix. ${\mathcal{G}} = \operatorname{diag} \left( {g_i} \right)$, where $g_i$ is the pinning gain from the leader to the $i^{th}$ converter. $g_i> 0$ if the leader links to the $i^{th}$ converter; otherwise, $g_i= 0$. ${{\mathbf{1}}_N} \in {\mathbb{R}^N}$ is a vector with all entries of one. $\left|  \cdot  \right|$ is the absolute value of a real number. $\operatorname{diag} \left\{  \cdot  \right\}$ constitutes a diagonal matrix from its set of elements.





For global voltage regulation and load sharing, the secondary control offers ${V_{n_i}}$ for each converter through data exchange with its neighbors. The dynamics of voltage droop and the secondary control can be described as
\begin{equation}
{\dot V}_{{n_i}} = \dot V_i^* + R_i^{\operatorname{vir} }{{\dot I}_i} = {{\bar u}_i} = {u_i} + {\delta _i},
\label{eq3}
\end{equation}
where ${V}_{{n_i}}$ is the reference for the primary control level, $V_i^*$ is the local voltage setpoint, $R_i^{\operatorname{vir} }$ is the virtual impedance, $I_i$ is the output current, ${{\bar u}_i}$ is the distorted input signal, $u_i$ is the control input to be designed, and ${\delta _i}$ denotes potential unbounded attacks on the input channels.

\textbf{\textit{Assumption} 1:} There exists a positive constant $\kappa_i$, such that $\left|{{\delta _i}\left( t \right)} \right|\le {\kappa_i} {t^\gamma }$.

For global voltage regulation and proportional load sharing under attacks, we employ an attack-resilient secondary control strategy at each converter based on neighborhood relative information. Denote ${\zeta_i} = {\sum\limits_{j \in {\mathcal{N}_i}} {{a_{ij}}\left( {{V_j} - {V_i}} \right)}  + {g_i}\left( {{V_{\operatorname{ref} }} - {V_i}} \right)}{ + \sum\limits_{j \in {\mathcal{N}_i}} {{a_{ij}}\left( {R_j^{\operatorname{vir} }{I_j} - R_i^{\operatorname{vir} }{I_i}} \right)} }$. We then present the following attack-resilient control protocols
 
\begin{equation}
\begin{array}{l}
  u_i = {\sum\nolimits_{\mu  = 0}^\gamma  {\xi _i^{\left( \mu  \right)}}}\Big( \sum\nolimits_{j \in {\mathcal{N}_i}} {a_{ij}}\left( {{V_j} - {V_i}} \right) 
  +{g_i}\left( V_{\operatorname{ref} } - V_i \right)\\+ \sum\nolimits_{j \in {\mathcal{N}_i}} {a_{ij}}\big( R_j^{\operatorname{vir} }{I_j}- R_i^{\operatorname{vir} }{I_i} \big)\Big) \\
  = \sum\nolimits_{\mu  = 0}^\gamma  {\xi _i^{\left( \mu  \right)}}\Bigg( \sum\nolimits_{j \in {\mathcal{N}_i}} {a_{ij}}\Big( \left( {{V_j} + R_j^{\operatorname{vir} }{I_j}} \right) - \left( {{V_i} + R_i^{\operatorname{vir} }{I_i}} \right) \Big)\\
  + {g_i}\Big(\big( V_{\operatorname{ref} } + R_i^{\operatorname{vir}}{I_i} \big) - \left( {{V_i} + R_i^{\operatorname{vir} }{I_i}} \right) \Big)\Bigg),
\end{array}
\label{eq5}
\end{equation}
where the term $\xi_i^{\left(\gamma\right)}$ in the coupling gain, is adaptively updated using the following tuning law
\begin{align}
&\xi_i^{\left(\gamma\right)} = \alpha_i\left({\zeta_i}^2-\upsilon_i\left(\xi_i^{\left(\gamma-1\right)}-\hat \xi_i^{\left(\gamma-1\right)}\right)\right), \label{eq7}\\
& \hat\xi_i^{\left(\nu\right)}  = \rho_i\left(\xi_i^{\left(\nu-1\right)}-\hat \xi_i^{\left(\nu-1\right)}\right), \label{eq8}
\end{align}
where $\alpha_i, \upsilon_i$ and $\rho_i$ are positive constants. In steady state, ${R_i^{\operatorname{vir} }{I_i}}$ converges to a constant $k I_{ss}^{\operatorname{pu} }$ \cite{7}. Denote ${\Theta _i} = {V_i} + {R_i^{\operatorname{vir} }{I_i}}$ and $\Theta _{\operatorname{ref} } = V_{\operatorname{ref} } + kI_{ss}^{\operatorname{pu} }$, we obtain

\begin{equation}
\begin{gathered}
  {{\dot \Theta }_i} = \sum\limits_{\mu  = 0}^\gamma {\xi _i^{\left( \mu  \right)}}\left( {\sum\limits_{j \in {\mathcal{N}_i}} {{a_{ij}}\left( {{\Theta _j} - {\Theta _i}} \right)}  + {g_i}\left( {\Theta _{\operatorname{ref} } - {\Theta _i}} \right)} \right)+{\delta _i}\hfill \\
\quad= \sum\limits_{\mu  = 0}^\gamma {\xi _i^{\left( \mu  \right)}} \left( { - \left( {{d_i} + {g_i}} \right){\Theta _i} + \sum\limits_{j \in {\mathcal{N}_i}} {{a_{ij}}{\Theta _j}}  + {g_i}\Theta _{\operatorname{ref} }} \right) + {\delta _i}. \hfill \\ 
\end{gathered}
\label{eq9}
\end{equation}

The global form of \eqref{eq9} is
\begin{equation}
\begin{array}{l}
\dot \Theta  =  - \operatorname{diag} \left(\sum\limits_{\mu  = 0}^\gamma  {\xi _i^{\left( \mu  \right)}} \right) \left( {\mathcal{L} + \mathcal{G}} \right)\left( {\Theta  -{{\mathbf{1}}_N}{\Theta _{\operatorname{ref} }} } \right)+\delta,
\end{array}
\label{eq10}
\end{equation}
where $\Theta = {[ {\Theta_1^T,...,\Theta_N^T} ]^T}$ and $\delta = {[ {\delta_1^T,...,\delta_N^T} ]^T}$ is the attack vector. Define the following global cooperative regulation error
\begin{equation}
\varepsilon  = \Theta  - {{\mathbf{1}}_N}{\Theta _{\operatorname{ref} }},
\label{eq11}
\end{equation}
where $\varepsilon = {[ {\varepsilon_1^T,...,\varepsilon_N^T} ]^T}$. Then, we obtain
\begin{equation}
\dot \varepsilon  =  - \operatorname{diag} \left(\sum\nolimits_{\mu  = 0}^\gamma  {\xi _i^{\left( \mu  \right)}} \right)\left( {\mathcal{L} + \mathcal{G}} \right)\varepsilon  + \delta .
\label{eq12}
\end{equation} 

As shown in \cite{am29}, we need to stabilize the cooperative regulation error $\varepsilon$ to achieve the global voltage regulation and proportional load sharing.


\textbf{\textit{Assumption} 2:} The digraph $\mathscr{G}$ includes a spanning tree, where the leader node is the root. 



\textbf{\textit{Definition} 2:} Signal $x(t)\in {\mathbb{R}}$ is UUB with the ultimate bound $b$, if there exist constants $b,c>0$, independent of ${t_0} \geq 0$, and for every $a \in \left( {0,c} \right)$, there exists $t_1 = t_1 \left( {a,b} \right) \geq 0$, independent of $t_0$, such that $\left| {x\left( {{t_0}} \right)} \right| \leq a \Rightarrow \left| {x\left( t \right)} \right| \leq b,\forall t \geq {t_0} + {t_1}$ \cite{39}.


\textbf{\textit{Definition} 3 (Attack-resilient Secondary Control Problem):} 
Under the generally unbounded attacks on local input channels described in \eqref{eq3}, design local cooperative control protocols for each converter such that, for all initial conditions, $\varepsilon $ in \eqref{eq11} is UUB. That is, the bounded global voltage regulation and proportional load sharing are achieved.

Next, we give the main result of solving the attack-resilient secondary control problem for DC microgrids.

\textbf{\textit{Theorem} 1:} Given Assumptions 1 and 2, under the unbounded attacks described in \eqref{eq3}, let the attack-resilient secondary control protocols consist of \eqref{eq5}, \eqref{eq7} and \eqref{eq8}, then the cooperative regulation error $\varepsilon$ in \eqref{eq11} is UUB. That is, the attack-resilient secondary control problem is solved. 

\textbf{\textit{Proof}:}
Define $\tilde{\xi}_i^{\left(\gamma-1\right)}\left(t\right)=\xi_i^{\left(\gamma-1\right)}\left(t\right)-\hat{\xi}_i^{\left(\gamma-1\right)}\left(t\right)$, then the derivative of $\tilde{\xi}_i^{(\gamma-1)}\left(t\right)$ is 
\begin{equation}
\begin{array}{l}
\tilde{\xi}_i^{\left(\gamma\right)}\left(t\right)=\xi_i^{\left(\gamma\right)}\left(t\right)-\hat{\xi}_i^{\left(\gamma\right)}\left(t\right)\\
\quad\;\;\,\,=\alpha_i\left({\zeta_i}^2\left(t\right)-\upsilon_i\left(\xi_i^{(\gamma-1)}\left(t\right)-\hat{\xi}_i^{(\gamma-1)}\left(t\right)\right)\right)-\\
\quad\;\;\,\,\rho_i\left(\xi_i^{(\gamma-1)}\left(t\right)-\hat{\xi}_i^{(\gamma-1)}\left(t\right)\right)\\
\quad\;\;\,\,=\alpha_i{\zeta_i}^2\left(t\right)-\left(\alpha_i\upsilon_i+\rho_i\right)\tilde{\xi}_i^{(\gamma-1)}\left(t\right).
\end{array} 
\label{eq14}
\end{equation}
\vspace{-2mm}
The solution of \eqref{eq14} can be written as
\begin{equation}
\begin{split}
\tilde{\xi}_i^{\left(\gamma-1\right)}\left(t\right)=&e^{-\left(\alpha_i\upsilon_i+\rho_i\right)t}\tilde{\xi}_i^{\left(\gamma-1\right)}\left(0\right)\\
&+\alpha_i\int_0^te^{-\left(\alpha_i\upsilon_i+\rho_i\right)\left(t-\tau\right)}{\zeta_i}^2\left(\tau\right)\operatorname{d}\tau.
\end{split}
\label{eq15}
\end{equation}

Since $e^{-\left(\alpha_i\upsilon_i+\rho_i\right)\left(t-\tau\right)}{\zeta_i}^2\left(\tau\right)$ is UUB, we obtain that $\tilde{\xi}_i^{(\gamma-1)}\left(t\right)$ is also UUB. According to Definition 2, let the ultimate bound of $\tilde{\xi}_i^{(\gamma-1)}\left(t\right)$ to be $\eta$. Note that the initial values of the gains are chosen such that $\tilde{\xi}_i^{(\gamma-1)}\left(0\right) \ge0$.

The global form of $\zeta_i\left(t\right)$ is
\begin{equation}
\zeta\left(t\right)=-\left(\mathcal{L} + \mathcal{G}\right)\varepsilon\left(t\right),
\label{eq16}
\end{equation}
where $\zeta\left(t\right) = {[ {\zeta_1^T\left(t\right),...,\zeta_N^T\left(t\right)} ]^T}$. Then, using \eqref{eq12} to obtain the time derivative of \eqref{eq16} as
\begin{equation}
\begin{array}{l}
\dot\zeta\left(t\right)=-\left(\mathcal{L} + \mathcal{G}\right)\left(\operatorname{diag} \left(\sum\nolimits_{\mu  = 0}^\gamma  {\xi _i^{\left( \mu  \right)}\left(t\right)} \right)\zeta\left(t\right) +\delta\left(t\right)\right),
\end{array}
\label{eq17}
\end{equation}

Define the following Lyapunov function candidate
\begin{equation}
E \left( t \right)= \frac{1}{2}\sum\limits_{i  =1}^{N} {\int_0^{{\zeta_i}^2\left(t\right)} { {\sum\limits_{\mu  = 0}^\gamma  {\xi _i^{\left( \mu  \right)}\left( s \right)} }} } {\mathop{\rm d}\nolimits} s . 
\label{eq18}
\end{equation}

The time derivative of \eqref{eq18} along the system trajectory of \eqref{eq17} is given by
\begin{equation}
\begin{array}{l}
\dot E \left( t \right)
=\frac{1}{2}\sum\nolimits_{i  = 1}^N {\sum\nolimits_{\mu  = 0}^\gamma  {\xi _i^{\left( \mu  \right)}\left(t\right)} 2{\zeta _i}{{\left( t \right)}}{{\dot \zeta}_i}\left( t \right)} \\
= \zeta {\left( t \right)^T}{{\mathop{\rm diag}\nolimits} \left( {\sum\nolimits_{\mu  = 0}^\gamma  {\xi _i^{\left( \mu  \right)}\left(t\right)} } \right)}\dot \zeta \left( t \right)\\
= -\zeta {\left( t \right)^T}{{\mathop{\rm diag}\nolimits} \left( {\sum\nolimits_{\mu  = 0}^\gamma  {\xi _i^{\left( \mu  \right)}\left(t\right)} } \right)}\\
\times\left(\mathcal{L} + \mathcal{G}\right)\Bigg(\operatorname{diag} \left(\sum\nolimits_{\mu  = 0}^\gamma  {\xi _i^{\left( \mu  \right)}\left(t\right)} \right)\zeta\left(t\right)+{\delta\left(t\right)}\Bigg)\\
\leq-\sigma_{\min}\left(\mathcal{L} + \mathcal{G}\right)\left\|\operatorname{diag} \left(\sum\nolimits_{\mu  = 0}^\gamma  {\xi _i^{\left( \mu  \right)}\left( t \right)} \right) \zeta\left(t\right)\right\|^2\\
+\sigma_{\max}\left(\mathcal{L} + \mathcal{G}\right)\left\|\operatorname{diag} \left(\sum\nolimits_{\mu  = 0}^\gamma  {\xi _i^{\left( \mu  \right)}\left( t \right)} \right) \zeta\left(t\right)\right\|\left\|\delta\left(t\right)\right\|\\
=-\sigma_{\min}\left(\mathcal{L} + \mathcal{G}\right)\left\|\operatorname{diag} \left(\sum\nolimits_{\mu  = 0}^\gamma  {\xi _i^{\left( \mu  \right)}\left( t \right)} \right) \zeta\left(t\right)\right\|\\
\left(\left\|\operatorname{diag} \left(\sum\nolimits_{\mu  = 0}^\gamma  {\xi _i^{\left( \mu  \right)}\left( t \right)} \right) \zeta\left(t\right)\right\|-\frac{\sigma_{\max}\left(\mathcal{L} + \mathcal{G}\right)\left\|\delta\left(t\right)\right\|}{\sigma_{\min}\left(\mathcal{L} + \mathcal{G}\right)}\right).
\end{array} 
\label{eq19}
\end{equation}

Given Assumption 2, $\left( {\mathcal{L} + \mathcal{G}} \right)$ is positive-definite \cite{38}. Denote $\beta=\frac{\sigma_{\max}\left(\mathcal{L} + \mathcal{G}\right)}{\sigma_{\min}\left(\mathcal{L} + \mathcal{G}\right)}$, which is a positive constant. Next, we will prove that
\begin{equation}
\begin{gathered}
\left\|\operatorname{diag} \left(\sum\nolimits_{\mu  = 0}^\gamma  {\xi _i^{\left( \mu  \right)}\left( t \right)} \right) \zeta\left(t\right)\right\|-\beta\left\|\delta\left(t\right)\right\|\geq0.
\end{gathered}
\label{eq20}
\end{equation}

Note that, a sufficient condition to guarantee (\ref{eq20}) is
\begin{equation}
\sum\nolimits_{\mu  = 0}^\gamma  {\xi _i^{\left( \mu  \right)}\left( t \right)}{\zeta_i}\left(t\right)\geq\beta\left|\delta_i\left(t\right)\right|.
\label{eq21}
\end{equation}

Pick $\left| \zeta_i\left(t\right) \right| \ge \sqrt{\mathop{\upsilon_i}\eta+\frac{\gamma!\kappa_i}{\alpha_i}}$, then based on Assumptions 1 and \eqref{eq5}, there exists $T\ge0$, such that $\sum\nolimits_{\mu  = 0}^\gamma  {\xi _i^{\left( \mu  \right)}\left( t \right)}\ge\left|\delta_i\left(t\right)\right|,\forall t\ge T$. Pick $\left| \zeta_i\left(t\right) \right|\ge \max \bigg\{\sqrt{\upsilon_i\eta+\frac{\gamma!\kappa_i}{\alpha_i}},\sqrt{\beta} \bigg\}$, then \eqref{eq21} and hence, \eqref{eq20} are guaranteed. Combining \eqref{eq19} and \eqref{eq20} yields
\begin{equation}
\begin{split}
\dot E\left( t \right) \le 0,\:\forall \left| \zeta_i\left(t\right) \right| \ge \max \bigg\{\sqrt{\upsilon_i\eta+\frac{\gamma!\kappa_i}{\alpha_i}},\sqrt{\beta} \bigg\},
t \ge T.
\end{split}
\label{eq22}
\end{equation}

From the LaSalle’s invariance principle, we obtain that $\zeta\left(t\right)$ is UUB. Finally, based on \eqref{eq16}, $\varepsilon\left(t\right)$ is also UUB. This completes the proof.
\hfill\(\blacksquare\)

\section{Validation and Results}

A low-voltage DC microgrid
is modeled to study the effectiveness of the proposed results. The converters have similar typologies but different ratings, i.e., the rated currents are equal to $I_i^{\operatorname{rated}}=(6,3,3,6)$, and virtual
impedance are equal to $R_i^{\operatorname{vir}}=(2,4,4,2)$. 
The converter parameters are $C = 2.2 \,\mathrm{mF}$, $L = 2.64 \,\mathrm{mH}$, $f_s = 60 \,\mathrm{kHz}$, $R_{line}=0.1\,\Omega$, $R_L = 20 \,\Omega$, $v_{ref} = 48 \,\mathrm{V}$, and $v_{in} = 80 \,\mathrm{V}$. 
The rated voltage of the DC microgrid is $48 \,\mathrm{V}$. Consider the FDI attack to the local control input of each converter by selecting $\delta_{i}=(0.8t^2+5,0.7t^2+5,0.8t^2+5,0.5t^2+5)$.

\begin{figure}[!h]
\centering
{\includegraphics[scale=0.4]
{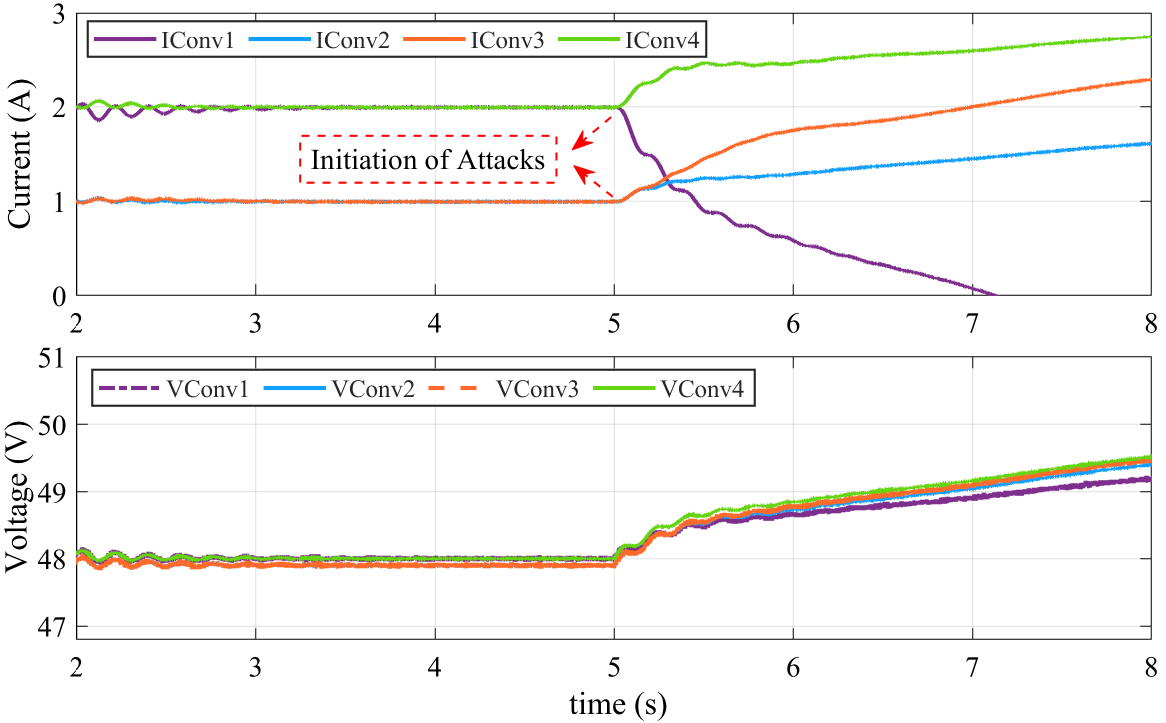}}
\caption { Performance of the resilient secondary control method in \cite{zhou2023distributed}.}
\label{FIG22}
\end{figure}

\begin{figure}[!h]
\centering
{\includegraphics[scale=0.4]{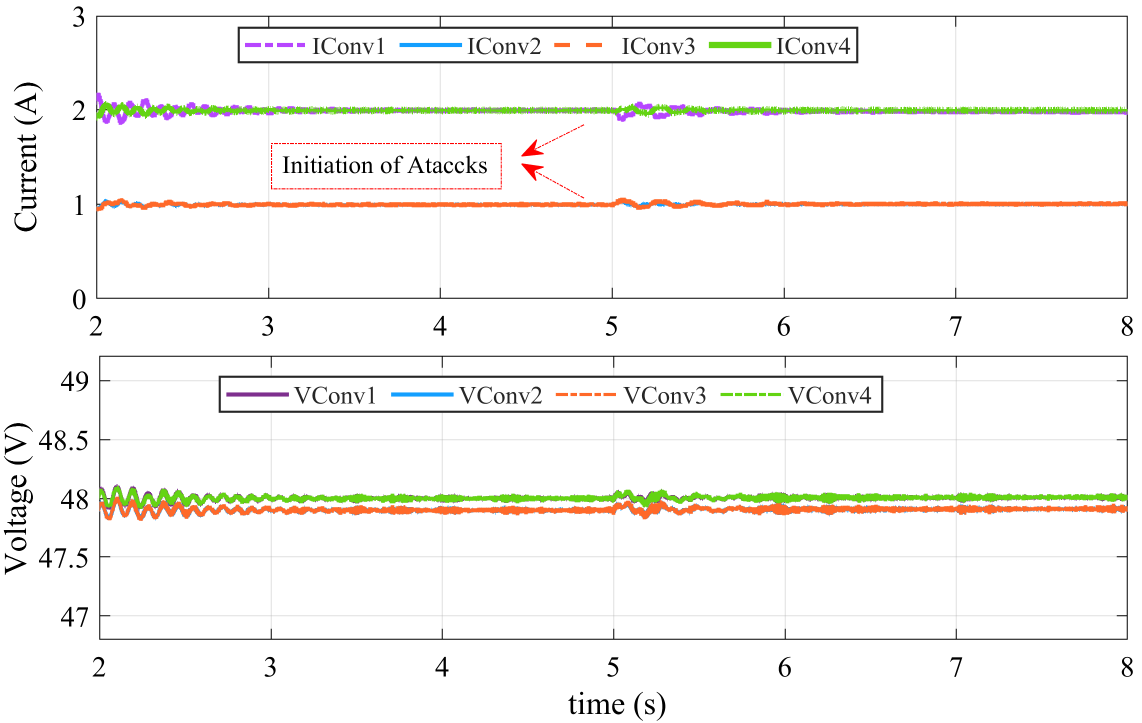}}
\caption {Performance of the proposed attack-resilient control method.}
\label{Fig23}
\end{figure}

Figures ~\ref{FIG22} and ~\ref{Fig23} compare the voltage and current responses against generally unbounded attacks using the resilient secondary control of \cite{zhou2023distributed} and the proposed resilient secondary defense strategy. The adaptive tuning parameters for the proposed resilient control method are set as  $\alpha_i = 1.5$, $\xi _i(0)=1$, $\dot \xi _i(0)= 70$, $\hat\xi _i(0)=1$, $i = 1, 2, 3, 4$. As seen, after initiating the attack injections at $t=5 \,\mathrm{s}$, both voltage and current diverge using the resilient method in \cite{zhou2023distributed}. In contrast, by utilizing the proposed attack-resilient secondary defense strategy, the voltage of each converter converges to a value within a small neighborhood of the reference value $48\, \mathrm{V}$, and the currents converge to values within small neighborhood around the two respective values reflecting the properly shared current. These validate that, while the existing resilient method fails to preserve the microgrid's stability in the face of generally unbounded attacks, the proposed resilient defense strategy successfully accomplishes the UUB convergence on both voltage regulation and current sharing for DC microgrids. 


\section{Conclusion}
This letter has proposed a novel secondary defense strategy for DC microgrids against generally unbounded FDI attacks. In contrast to existing resilient methods that impose strict requirements for bounding the first time derivatives of unbounded attack signals, the proposed defense strategy relaxes such a stringent constraint by addressing a wider range of unbounded attack signals, significantly bolstering the cyber-physical resilience of DC microgrids. Rigorous proof, based on Lyapunov techniques, has shown that the proposed strategy guarantees the UUB convergence for both voltage regulation and proportional load sharing under generally unbounded attacks. Comparative case studies have validated the enhanced resilience of the proposed control method.



\ifCLASSOPTIONcaptionsoff
  \newpage
\fi
\bibliographystyle{IEEEtran}

\bibliography{Ref_DC}

\end{document}